# A Cloud-Based Tool for Meteorite Recovery Using Drones and Machine Learning

Seamus L. Anderson[1, 2], Hadrien A. R. Devillepoix[2, 3], Lewis Lakerink[4], Sawitchaya Tippaya[5], Dale P. Giancono[2, 3], Martin C. Towner[2], Iona Clemente[2, 3], Martin Cupák[2], Ashley F. Rogers[2], John H. Fairweather[6], Mia Walker[2], Daniel Burgin[2], Michael A. Frazer[2, 3], Sophie E. Deam[2, 3], Veronika Pazderová[7], Eleanor K. Sansom[3], Benjamin A. D. Hartig[3], Hely C. Branco[2], Thomas Stevenson[2, 3], Isabella Hatty[2, 3], Anna Zappatini[8], Anthony Lagain[2, 9], Tom Lovelock[10], Auriane Egal[11, 12], Lucy Forman[2], David Belton[13], Simon Windsor[2], Shibli Saleheen[4], Asher Leslie[4], Gregory B. Poole[4], Andrew Langendam[14], Rachel S. Kirby[2, 15], and Andrew G. Tomkins[15]

*[1]NASA Goddard Space Flight Center, Greenbelt, MD 20771, United States.*
*[2]Space Science and Technology Centre, Curtin University, GPO Box U1987, Perth WA 6102, Australia.*
*[3]International Centre for Radio Astronomy Research, Curtin University, GPO Box U1987, Perth WA 6102, Australia.*
*[4]Astronomy Data and Computing Services (ADACS), The Centre for Astrophysics & Supercomputing, Swinburne University of Technology, P.O. Box 218, Hawthorn, VIC 3122, Australia.*
*[5]Curtin Institute for Data Science, Curtin University, GPO Box U1987, Perth WA 6102, Australia.*
*[6]Centre for Rock Art Research and Management, University of Western Australia, Perth WA 6009, Australia.*
*[7]Faculty of Mathematics, Physics and Informatics, Comenius University Bratislava, Mlynská dolina, 84248, Bratislava, Slovakia.*
*[8]Institute of Geology, University of Bern, Bern, Switzerland.*
*[9]Aix-Marseille University, CNRS, IRD, INRA, CEREGE, Institut Origines, Aix en Provence, France.*
*[10]Royal Holloway University of London, Surrey, TW20 0EX UK.*
*[11]Planétarium de Montréal, Espace pour la Vie, 4801 av. Pierre-de Coubertin, Montréal, H1V 3V4, Québec, Canada.*
*[12]Department of Physics and Astronomy, The University of Western Ontario, London, Ontario, Canada.*
*[13]School of Earth and Planetary Sciences, Curtin University, GPO Box U1987, Perth WA 6102, Australia.*
*[14]Australian Nuclear Science and Technology Organisation, Clayton, Victoria 3168, Australia.*
*[15]School of Earth, Atmosphere and Environment, Monash University, Melbourne, Victoria 3800, Australia.*

**ABSTRACT**

We present a cloud-based tool that uses drones and machine learning to help recover instrumentally observed meteorite falls. We showcase a collection of improvements made upon previous iterations of our system, as well as detail the successes and limitations of this technique when applied to observed meteorite falls in South and Western Australia. This tool is available to the meteoritics research community upon request at https://find.gfo.rocks.

**INTRODUCTION**

Simply described, meteorites are rocks that fall from the sky and can originate from a number of rocky celestial bodies including Mars, Luna, and a great number of asteroids. For the vast majority of the more than 78,000 meteorites in the global collection, there is no spatial context as to how they arrived on Earth. The Příbram meteorite fall (Ceplecha, 1961) marked the first instance where a fireball event was photographically observed well enough to calculate a prior orbit, estimate the area on the ground where surviving pieces may have fallen, and finally recover the meteorite for geologic study. Thus began the era of fireball networks. At present, over a dozen fireball and meteor observation networks are maintained by a number of scientists and institutions around the globe (Bland et al., 2012; Devillepoix et al., 2020; Colas et al., 2020; Oberst et al., 1998; Trigo-Rodríguez et al., 2006; Gardiol et al., 2021; Cooke et al., 2012; Gritsevich et al., 2014; Brown et al., 2011}, with the goal of characterizing the meteoroid environment in near Earth space, as well recover meteorites for which their orbits prior to impacting the Earth are constrained. More recently, satellite-based photographic sensors and Doppler weather radar stations are being used in concert with, and sometimes independently of these fireball networks to expand the area of observation, and therefore the chance of instrumentally observing and recovering a meteorite fall (Fries and Fries 2010; Fries and Fries 2014; Fries et al., 2025; Jenniskens et al., 2012; Devillepoix et al., 2022a).

When these well-observed meteorites fall near population centers, the financial and labor costs of recovering the meteorite are often significantly reduced or eliminated almost entirely. For instance, the Winchcombe meteorite, which fell in the UK on the night of 28th February 2021, was first located the following morning by a family who noticed that it had impacted their driveway (King et al., 2022; Russell et al., 2024), with many other pieces being recovered by the ensuing searches on foot later that week. Similarly, a fragment of the Ejby meteorite (which fell in a suburb of Copenhagen) was found by a member of the public after they noticed the unusual rock on the tiles outside their front door (Haack et al., 2019). Possibly the most harrowing recovery of an orbital meteorite was that of the Golden meteorite, which fell through the roof of a private home in Canada, landing on a pillow next to a person who was asleep in bed just moments before (Brown et al., 2023). The meteorites Ribbeck (Spurný et al., 2024) and Saint-Pierre-le-Viger (Bischoff et al., 2023; Zanda et al., 2023; Egal et al., 2025) were each recovered less than 1 km from the town centers of their namesakes, by teams of more than a dozen people. Falls close to population centers such as these increase the pools of possible volunteer searchers, which often include members of the public who live nearby and whose costs only include time, as they can return home each night, and require little specialized equipment beyond adequate outdoor clothing. It is also not uncommon for planetary scientists or private meteorite hunters who live nearby, or less than one day's travel away, to volunteer or spend their time searching.

As the distances between observed meteorite falls and significantly populated communities increase, the costs to search for said meteorites also increase. Nearly every orbital meteorite recovered via the Desert Fireball Network (Spurný et al., 2012; Spurný et al., 2012b; Sansom et al., 2020; Devillepoix et al., 2018; Shober et al., 2022; Devillepoix et al., 2022b, Anderson et al., 2022), which operates in South and Western Australia (Bland et al., 2012; Howie et al., 2017), fell more than a day's drive away from a population center that hosts an airport with regularly scheduled flights. Often the final portion of the more than two-day journey to these meteorite fall sites includes a drive of more than 50 km on unpaved roads or no roads at all (typically more than 2 hours in a 4-wheel-drive vehicle). As a result, these remote fall sites often necessitate on-site camping as the most efficient plan of action, which in turn adds to the material requirements of the trip. These remote trips also present a heightened medical risk to those involved. Even though a wound from a dangerous animal, a broken leg from a bad step, heat stroke, a heart attack, anaphylaxis, and most other medical emergencies, could conceivably happen during any searching trip including those close to a populated area, the time between an injury occurring and the patient receiving adequate medical care significantly increases with the remoteness of these distant fall sites. Contingency plans to treat such medical emergencies usually require first-aid training for all party members, a first aid kit, an Automated External Defibrillator, a medical airlift, and redundant communication systems in the form of satellite phones and satellite internet, all of which further add to the material requirements of the trip. For these reasons, limiting both the number of people and the number of days needed to search for extremely remote meteorite falls is a critical objective for both cost and more importantly, safety. The most obvious and successful approach to this problem employs the use of drones and machine learning.

A range of previous works have explored the feasibility of using drones and/or machine learning to recover meteorites in a few distinct ways including hyperspectral imaging, magnetic sensors, thermal imaging, and optical imaging. Regardless of the method, any attempt to use drones and signal processing to recover meteorites must balance the following constraints: survey rate, data processing rate, detection limit (meteorite size), true positive detection rate, false positive detection rate, and hardware cost. For instance, most meteorite fall sites can be surveyed in less than one day if the drone flies at maximum altitude but will likely fail to detect all but the very largest meteorites. Conversely, if the drone is flown very close to the ground, it is possible to detect any meteorite that a searcher on foot is capable of identifying, but the survey would likely take weeks, with the data processing taking even longer, and would likely suffer from an unmanageable number of false positives. With these constraints in mind, some methods have proven more practical than others.

Hyperspectral imaging was explored by Moorhouse, 2014 who showed that meteorite interiors had unique spectral signatures that could be recognized via machine learning. A possible limitation of this approach is the lack of extension to meteorite exteriors, particularly those with fusion crusts endemic to orbital meteorite falls. Another potential drawback is the comparatively low spatial resolution of any hyperspectral camera, which would dramatically inflate the time required to survey a fall site, since the drone would have to fly at a lower altitude to achieve the same spatial resolution. This closer proximity to the ground would also increase the chance of crashing into obstacles. A hyperspectral camera is also specialized equipment and not likely to be as available or as affordable as a traditional RGB camera.

Magnetic sensing of meteorites was initially studied by Su, 2017, who showed the feasibility of this approach, which could conceivably detect the majority of meteorites. An issue with this detection method involves the scientific value of non-magnetic meteorites that comprise the minority. Although ordinary chondrites and iron meteorites would both display a detectable magnetic signature, some of the rarest, and therefore, scientifically valuable meteorites do not contain magnetic mineral phases, meaning that many carbonaceous chondrites and achondrites would be passed over by this approach. Another issue with this methodology is the required survey altitude of less than a few meters. Besides inflating the time required to survey the fall site, this low altitude would also increase the likelihood of colliding with ground-based obstacles. A final issue with this approach involves false positives from anthropomorphic objects. Although we occasionally find human-made metal objects such as nails, chains, and cans in the Nullarbor plain, other searching areas such as Europe and North America that are far more densely populated would encounter these false positives far more frequently.

Thermal imaging was most extensively investigated by Hill et al., 2023, who showed the unique advantage of using drone-mounted thermal cameras to search for meteorites in snow-covered strewn fields. Though currently, thermal sensors have a spatial resolution that is 3-4 orders of magnitude lower than a typical 20-40 MP RGB camera, likely requiring a much lower survey height to meaningfully resolve a meteorite. Although surveying fall sites in hot deserts using a thermal camera could leverage the unique thermal signatures presented by black, fusion-crusted meteorites in hot environments (Al-Owais et al., 2019), much work remains before this method is fully realized (Lovelock et al., 2025).

Striking a key balance between financial cost, survey speed, and image resolution, is the approach of using an RGB camera to survey the fall site, and a machine learning algorithm to identify meteorites in the images (Citron et al., 2017, Al-Owais et al., 2019; Anderson et al., 2020; Pons Recasens 2020, Citron et al., 2021; Anderson et al., 2022; Anderson et al., 2023}. The first documented effort to use a drone and machine learning for meteorite recovery was presented by Citron et al., 2017, who were also first to identify the problem posed by false positives. Zender et al., 2018 successfully used a color matching algorithm to identify test meteorites. Unfortunately, they encountered a significant number of background-dependent false positives, and at the time of writing, this effort has not been continued further. Al-Owais et al., 2019 and Al-Owais and Al-Khalifa 2026, present the only efforts so far to process the image data on-board the drone, an approach that will likely progress even further as edge-case computing and machine learning models advance. The efforts by both Citron et al., 2021 and Anderson et al., 2020 process the data post-flight using machine learning algorithms, which currently presents the most robust avenue for eliminating false positives, due to current limitations of machine learning model performance.

The previous work by Anderson et al., 2020 and Anderson et al., 2022 present a comprehensive strategy to eliminate false positives post-flight and investigate possible meteorite candidates in the field, with the latter efforts resulting in the first-ever recovery of an orbital meteorite fall using drones and machine learning. This methodology was later applied to a strewn field in South Australia that was discovered nearly 10 years after the event by analyzing data from satellite sensors and weather radar data (Devillepoix et al., 2022). In this strewn field more than 100 meteorite fragments were recovered, two of which were the result of drones and machine learning (Anderson et al., 2023). In this publication, we present a streamlined and publicly accessible tool for recovering orbital meteorites using drones and machine learning.

## METHODS

Since our early successes, we have made several improvements to our methodology including, the machine learning software, cloud processing of the data, crowdsourcing the meteorite candidate review, georeferencing meteorite candidates and the survey data, as well as a number of more subtle technical improvements to streamline the process. The upgraded system is hosted as a web app that can be found at https://find.gfo.rocks.

Although our process for using drones and machine learning to recover meteorites has been detailed in previous publications (Anderson et al., 2020; Anderson et al., 2022; Anderson, 2023), we briefly review the steps below for clarity:

- Multiple drone flights (approximately 1 $km^2$ each) are planned to image the fall site using an RGB camera at ~2 mm $pixel^{-1}$ (training data is also captured at this resolution)
- Training data of the background is gathered by flying a drone along the perimeter of the fall site and capturing an image every ~50 m, with the intent to sample a variety of background terrains and features present in the search area
- Training data of the meteorites is gathered by placing 20 or more meteorites (or black painted rocks) in a line ~5 m apart. Each example is imaged once by the drone camera as a person walks ~5 m next to the line, pointing at each one as they pass by. The images are labeled by drawing a rectangular box around each meteorite and its shadow
- A machine learning model is trained to identify meteorites while ignoring background features and is then applied to the survey images. The resulting meteorite candidates are passed to the user for four stages of false positive elimination

- Stage 1: Meteorite candidates are presented to the user in a 3x3 grid whereby they can quickly select promising candidates for further evaluation
- Stage 2: Remaining meteorite candidates are presented for elimination to the user one at a time, with the ability to adjust the magnification and inspect the entire image
- Stage 3: A follow up drone flies to visit the remaining meteorite candidates for further elimination
- Stage 4: The final meteorite candidates are visited on foot by the searchers until the meteorite is found or the list of candidates is exhausted

**Migration to a Cloud-Hosted Web Server**

Previously, all of the data collection, curation, labeling, and processing, as well as meteorite candidate sorting was completed onsite using a GPU-equipped desktop computer that demanded approximately 900 W of continuously supplied power (Anderson et al., 2020; Anderson et al., 2022). This computer was powered down every 2.5 hours to refuel the on-site generator, and was occasionally prone to failures involving heat, dust, moisture, and percussive events on the journey to the fall site. For these reasons, and in an effort to save space and weight when loading the vehicle, we have created a web-hosted application that performs all of the required tasks remotely (Fig. 1), such that now a searching team only needs one laptop to transmit the data, though multiple laptops are ideal for redundancy.

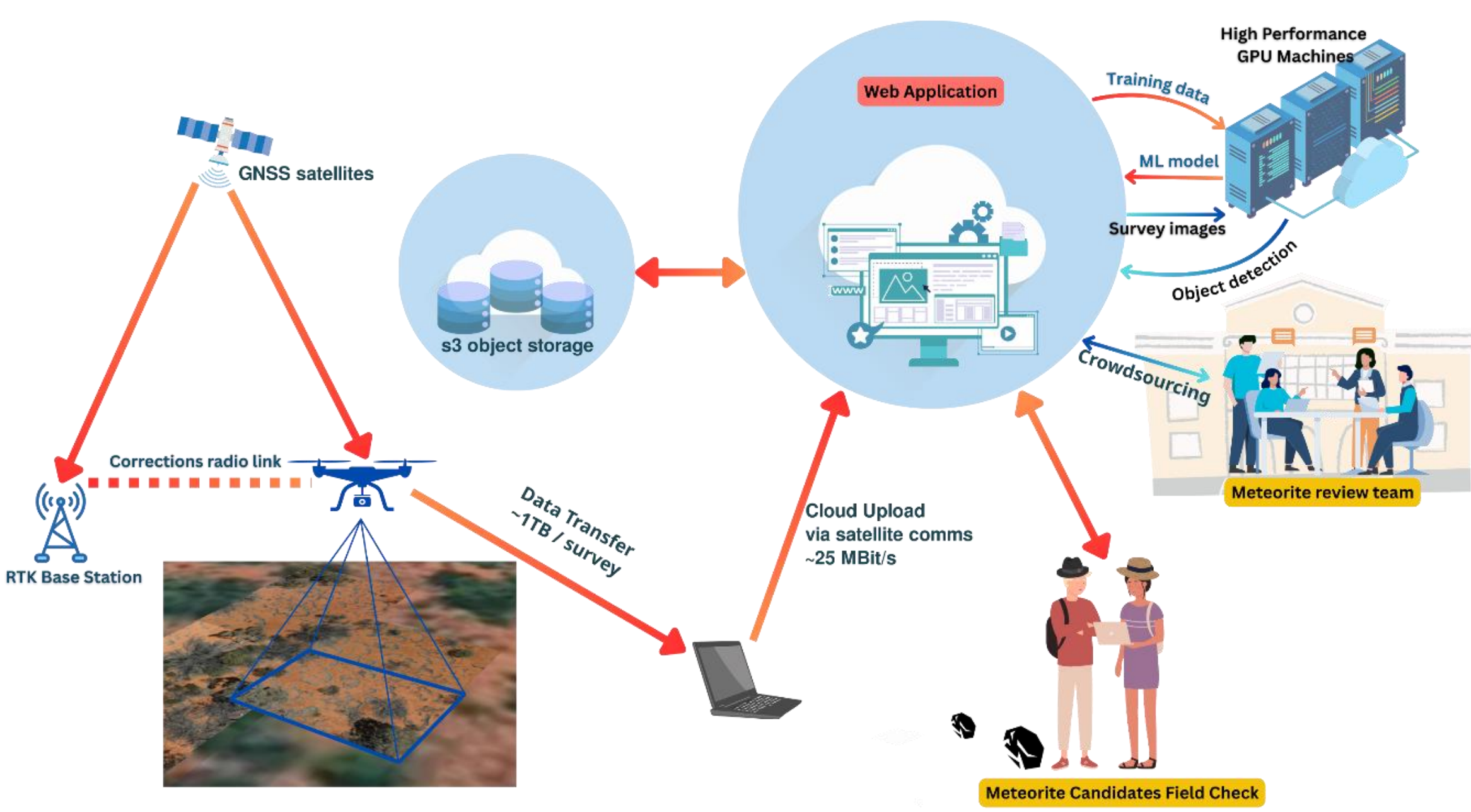

Fig. 1. Data processing overview: RGB imaging capture, data upload, data processing, and candidates human review process.

Using a mobile satellite communications terminal, ~30 Mb $s^{-1}$ can be continuously uploaded. On a typical day, our drone surveying system (a DJI M300 equipped with a Zenmuse

P1 48 MP camera) captures ~0.4 TB of JPEG RGB images. Although the satellite-upload process runs continuously even when the drone is not surveying, only ~81% of all of data captured can be transmitted over the course of a 24 h cycle. Though since the areas of higher meteorite probability are surveyed first, this is of little operational consequence, since by the end of the drone survey the probability of the meteorite image having been uploaded and processed would gradually increase beyond this 81% upload rate. By eliminating the requirement to operate a high-performance computer in the field, we have also reduced the on-site power requirements. The on-site laptop and satellite communications terminal are powered via a 2 kWh Lithium-Iron Phosphate battery which is recharged during the day using a solar panel, limiting generator usage to only the daylight hours when the drone batteries must be recharged.

The most significant advantage afforded by processing the data on a web platform is widening the bottleneck of meteorite candidate sorting. Previously, the meteorite candidate review process in Stages 1 and 2 could only be completed on one computer with one user at a time. Now that the candidate sorting interfaces are served by a web application, there is no limit to the number of users who can sort through meteorite candidates at any given time. This also redistributes the workload to team members who are not in the field, so that on-site searchers can focus their efforts on tasks that cannot be done remotely (e.g. flying the drone, eliminating stage 3 and 4 candidates), further reducing the on-site time required.

**High Fidelity Image Capture**

When collecting images for a meteorite-searching drone survey, a significant overlap is wastefully redundant, as it increases the total amount of data to be uploaded to, and processed by computing resources. It also increases the number of duplicate candidates to be reviewed by searching team members. Nevertheless, some overlap is necessary for preventing any gaps in the surveyed area to account for variations in terrain elevation, possible offsets in the drone's position, and imperfect pointing of the camera at nadir.

The drone's position is internally controlled by a GNSS system, which is notoriously imprecise with respect to altitude. We now supplement that system with a Real-Time Kinematics (RTK) base station to improve the accuracy of the drone's positioning. This enables us to set the frontal and lateral overlaps to 12% when planning each drone survey, as opposed to 20% previously. This reduction in duplicate data capture was a significant step in limiting the amount of data that must be uploaded to the cloud for processing. Although potential drawbacks to this overlap reduction include a lower quality orthorectification and a decrease in resolution along the image sides, we have not encountered these issues with our drone and camera, though systems with a lower camera resolution may suffer from this problem more acutely. The RTK addition has also noticeably improved the drone's ability to maintain a constant altitude and therefore constant height with respect to the ground, which is mostly flat in the typical areas we search on the Nullarbor Plain. It has also dramatically improved the efficiency of in-situ follow visits in Stage 4.

Finally, areas of higher meteorite probability (following the dark flight calculations presented by Towner et al., 2022) are surveyed and processed first, advancing the time when the survey can be interrupted because the meteorite has been located.

**Improved Machine Learning Engine**

The process of producing training data remains mostly the same as before, though the web-hosted software now has built-in functionality to allow users to label meteorite training data, rather than using a separate software program such as ImageJ (Schneider et al., 2012) or Label Studio (https://github.com/heartexlabs/label-studio).

A significant change between the current software and the approach described in Anderson et al., 2022 is the adopted use of the YOLOv8 algorithm (Jocher et al., 2023) as opposed to the previously used Keras engine (Chollet, 2015). Our previous approach would slice a 48 MP image into ~10,000 overlapping tiles, then perform image recognition on these tiles to determine if each one was an image of a meteorite. Any tile that had a confidence score greater than 0.7 was reported to the user. As described in Anderson et al., 2022, the imbalance in training data between background tiles and meteorite tiles was approximately 40:1, and required a specialized training routine, that involved changing the background tiles during training, keeping the meteorite tiles fixed, and maintaining a ratio at 1:1 between the two training sets. This has been completely supplanted with the YOLOv8 object detection engine, which addresses this fundamental limitation of a tile-based classification approach.

The meteorite searching dataset that we use to train each model is currently comprised of images from 13 separate survey campaigns in Western and South Australia, captured across three different camera systems: a DJI Mavic Pro (12 MP), a Sony α 7R Mk.III (42 MP), and a Zenmuse P1 (48 MP). Images from each searching trip are typically added to this compiled dataset for use in future trips. Currently, the dataset encompasses 751 images in total (429 positive samples with meteorite annotations (57.1%) and 322 background-only images (42.9%)), with 60% being split into training, 20% into validation, and 20% into the test set. Each full-resolution image was sliced into 320 × 320-pixel tiles with a 70-pixel overlap in both x and y. We discarded tiles where the annotated meteorite was partially cutoff as well as partial boundary tiles (dimensions < 320 × 320 pixels) lacking meteorite annotations. We strided over each meteorite location such that the meteorite would appear in one of 81 locations within the 320 × 320-pixel tile. This generated supplementary tiles that captured diverse viewpoints of each meteorite. The complete tile extraction process generated 525,826 individual tiles across all subsets of the total 751 original source images. For each survey location, we add locally acquired positive (meteorite) and negative (background) images, and train a new model from non-initialized weights, without transfer learning from COCO (Common Objects in Context) pre-trained weights, due to substantial differences between natural images (COCO distribution) and these geological survey images. We also conducted an experiment where we trained three models using uninitialized weights for 100 epochs, with identical parameters (see Table 1) across three different training set balances to examine model performance across different class balances:

- Baseline Configuration: Retention of all tiles preserving natural class distribution (94:6 background-to-meteorite ratio)
- Reduced Background Configuration: Random subsampling of background tiles to 50% of training data (approximate 1:1 balance)
- Meteorite-Only Configuration: Exclusive retention of positive samples, eliminating all background tiles

Table 1. The hyperparameters we use to train the YOLOv8-small model

| Parameter | Value | Rationale |
|---|---|---|
| Batch Size | 128 | Maximizes GPU memory utilization |
| Image Size | 320 × 320 | Matches tile extraction geometry |
| Optimizer | Adam W (default) | Adaptive learning rate with weight decay |
| Learning Rate Schedule | Cosine annealing | Smooth decay towards convergence |
| Mosaic Augmentation | Disabled | Objects never overlap |
| Scale Augmentation | Disabled | Preserves target scale in survey data |
| Flip Augmentation | Horizontal & Vertical | Appropriate for rotationally symmetric objects |
| Mixup | Disabled | Inappropriate for geological target detection |
| Early Stopping Patience | 5-80 epochs | Terminates training when validation loss plateaus |

We evaluated each model's performance using standard object detection metrics, based on the number of true positives (TP), false positives (FP), and false negatives (FN):

- Precision: $P = TP / (TP + FP)$; proportion of positive predictions that are correct
- Recall: $R = TP / (TP + FN)$; proportion of actual positive instances correctly identified
- F1-Score: $F1 = 2 \cdot PR / (P + R)$; harmonic mean of precision and recall
- mAP50: Mean Average Precision at 50 % Intersection-over-Union (IoU) threshold
- mAP50-95: Mean Average Precision averaged across IoU thresholds 50 % to 95 %; more stringent localization requirement

The results of the training experiment are shown in Table 2 and suggest a few notable findings. The mAP50-95 stability (0.6076 to 0.6311) across all configurations suggests that class balance (meteorite vs. background) primarily affects classification performance (precision / recall) rather than localization (bounding box) accuracy. The consistency between validation and test metrics across all configurations indicates good generalization with minimal overfitting. The slightly higher recall in the meteorite-only configuration (with respect to the baseline) suggests that it is marginally better for maximizing the chance of positively identifying a meteorite. The persistent high performance across all configurations (F1-scores 0.9618-0.9752) indicates robust meteorite detection regardless of class balance. This high performance also suggests that further improvements may come from model architecture innovations or expanded training datasets, rather than class balance fine tuning.

Table 2. Validation and Test set results for training on three different data configurations.

| Configuration | Epochs | Validation Precision | Validation Recall | Test Precision | Test Recall | mAP50 | mAP50-95 | F1-Score |
|---|---|---|---|---|---|---|---|---|
| Baseline | 30 | 0.9538 | 0.9344 | 0.9968 | 0.9391 | 0.9543 | 0.6151 | 0.9671 |
| Even Split | 31 | 0.9694 | 0.9446 | 0.9902 | 0.9350 | 0.9677 | 0.6076 | 0.9618 |
| Meteorite Only | 51 | 0.9908 | 0.9638 | 0.9909 | 0.9600 | 0.9880 | 0.6311 | 0.9752 |

Our previous keras-based approach typically performed with effectively the same test precision shown here, 0.9901, and slightly worse test recall of 0.9100 indicating an improvement correctly identifying meteorites. In practice though, we have seen an improvement in false positive suppression, where the keras approach produced approximately 17 meteorite candidates per image, the YOLOv8 engine typically produced between 1.8 and 2.8 meteorite candidates per image.

**Stage 4 Georeferencing**

In our previous methodology, the Stage 3 meteorite candidate review process involved sending a follow-up drone to take a closer look at the candidate and determine if it could still plausibly be a meteorite. This feature has not yet been implemented in the new web app, and therefore users proceed directly from Stage 2 to Stage 4. A current compromise allows field searchers to view the stage 4 candidate before they walk to it, allowing them to eliminate it before visiting in-person. This is possible because of knowledge and intuition of the local area gained by the field searchers after visiting a number of candidates, whereby they may be able to recognize obvious false positives that would not appear obvious to teleworking searchers who are not in the field.

Stage 4 is the final check for the remaining meteorite candidates and is performed by a team member in the field physically walking to each of their locations. This process was somewhat cumbersome, requiring a physical file-upload to a mobile device, the use of multiple applications to navigate to the meteorite candidate, and a separate built-in file browser to view the image, all while cross-referencing image filenames with GPS coordinates. The user then had to mentally orient themselves to register the image relative to their local background in order to find the candidate. Stage 4 is now integrated into the web platform as an interactive map (as shown in Fig. 2) that users can access on a mobile device. Although a smartphone can be used, the larger screen of a tablet allows for a better user experience. The geo-referenced drone survey image that contains the meteorite candidate is overlaid on top of satellite imagery, upon which Stage 4 candidates appear as blue icons. The user's live location and compass bearing are also displayed on the map to enable easier navigation to the meteorite candidates. We have found that geo-referencing the survey images captured by an RTK-enabled drone is accurate enough to guide the user to within 1-3 m absolute location of the meteorite candidate, which more or less corresponds to the GNSS precision achieved by a mobile device in open terrain. Upon inspection of each candidate, the user can either accept it as a meteorite and begin sample acquisition procedures or reject it with a possibility of assigning a class of false positive.

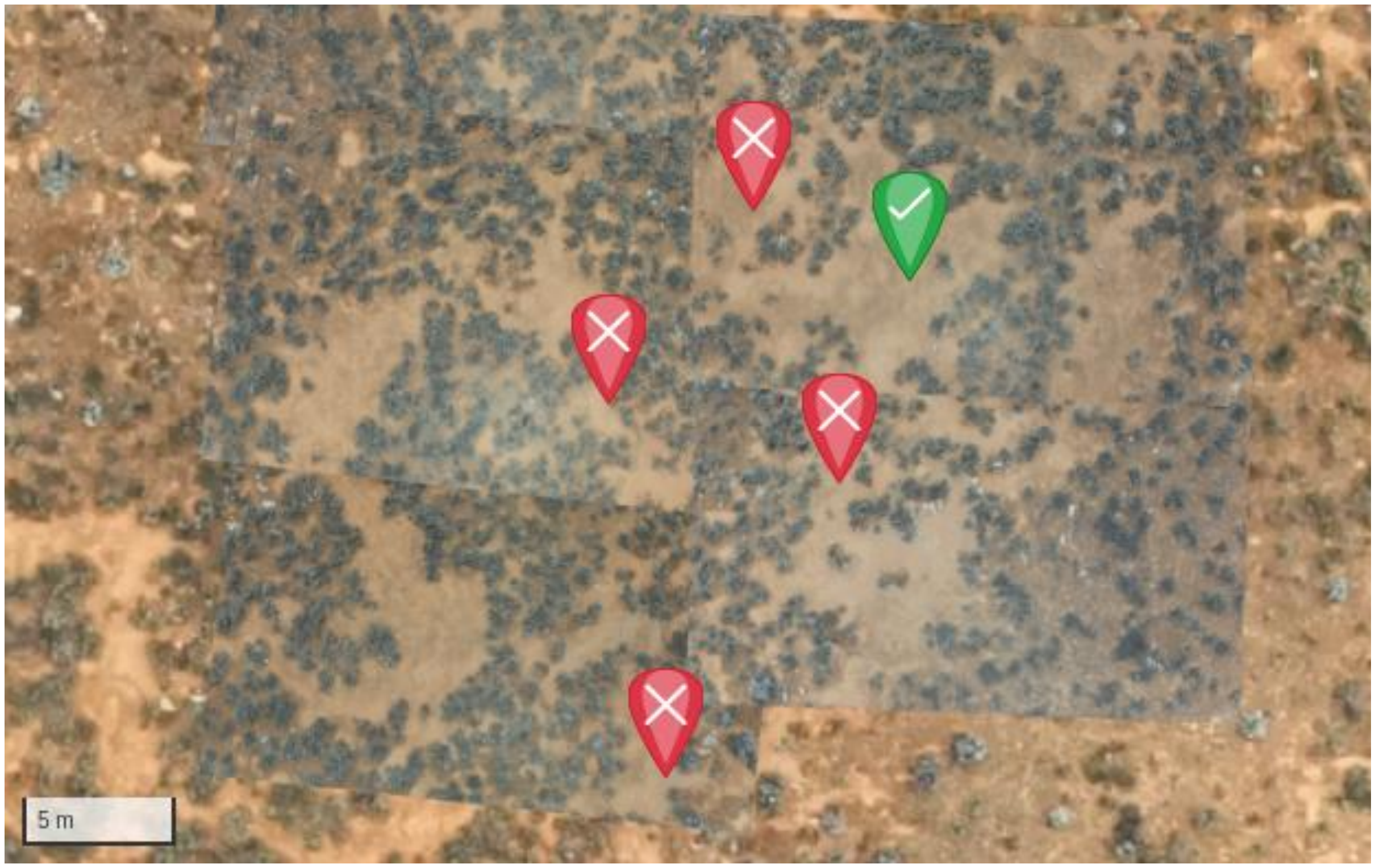

Fig. 2. A screenshot of the Stage 4 function in the webapp at the DN230523_02 fall area. The background is an orthomosaic generated from low-resolution drone data (100 mm pixel$^{-1}$). High-resolution survey images (2 mm pixel$^{-1}$) that contain stage 4 meteorite candidates, georeferenced and overlaid as aid to the user. Here the operators have visited the 5 targets, rejecting 4 of them (red markers), with the last one being a meteorite (green marker). Users are able to view the survey image of the meteorite candidate by selecting the marker.

## RESULTS AND DISCUSSION

### Efficiency of Stages 3 & 4

In later sections we outline the obstacles and opportunities facing the implementation of Stage 3 searching in the web application. This includes the addition of more sophisticated, but still reasonably accessible instruments such as thermal and multispectral drone-mounted cameras (Lovelock et al., 2025). As mentioned previously, we have significantly improved the user experience of Stage 4. Consequently, Stage 4 has become significantly more efficient, with one operator typically able to visit approximately 80 targets per 10-hour day. The streamlined experience also enables a previously mentioned temporary compromise while Stage 3 drone follow up searching is integrated into the web application. On-site searchers are encouraged to review their next targeted Stage 4 candidate before traveling there on foot, allowing them to discard obvious false positives that would only be obvious to on-site searchers with local site knowledge. We estimate that ~20-30% of Stage 4 candidates can be eliminated in this way.

## Meteorite Recovery Attempts: Successes and Failures in Australia

### Success: Kybo-Lintos

Our first successful recovery of a meteorite using drones and machine learning took place on Kybo Station in the Lintos paddock and was documented by Anderson et al., 2022 (see Fig. 3). This fall site was also the first instance where we were able to survey the majority of the fall area, with the meteorite being found after processing only ~8 % of the drone data. This early success significantly helped in securing the necessary funding to enable the development of the methodology presented in this work.

### Success: Benghazi Dam Strewn field

Our second, though limited, success occurred at another fall site where the fireball event was observed not through the Desert Fireball Network, but through a combination of Doppler radar data and satellite observations (Devillepoix et al., 2022; Anderson et al., 2023). Although this drone search did result in the recovery of two specimens (see Fig. 3) of the same orbital meteorite (and an additional ~100 specimens through traditional searching methods), multiple equipment failures and logistical bottlenecks spurred the work presented in this publication. For instance, our field computer (GPU-equipped desktop) experienced a critical failure on the first day that eventually required a new motherboard after returning from the field, highlighting the vulnerability of relying on on-site high-capacity computing resources. This fall site was also outside of our typical searching region of the Nullarbor plain, and contained more significant elevation changes as well as tree canopy heights often in excess of 20 m. Because of these more significant obstacles as well as changing weather patterns that brought them varying air pressure, the drone nearly collided (< 1 m clearance) with ground obstacles on three separate occasions. With the addition of the RTK ground station in this work, we have not since encountered any near collisions with ground obstacles.

### DN250711_02 Orbital Meteorite Recovery

The third success using drones and machine learning to recover an orbital meteorite fall, and the first success using this updated methodology via the web application and YOLOv8, occurred when searching for a meteorite that fell on the night of 11 July 2025, North-West of the Nullarbor plain. In this survey, conducted in November 2025, 10,818 images of the fall area were captured, covering 1.6 km$^2$, which encompassed the entirety of the calculated 2σ fall area. This resulted in 31,153 Stage 1 candidates, 2,159 Stage 2 candidates, and 728 Stage 4 candidates. This represents an average of 2.8 meteorite candidates identified by YOLOv8 per image, as opposed to the previously encountered average of 17.6 using keras for the Kybo-Lintos site, a significant improvement. During the candidate-false positive sorting stage of this 4-day trip, the three onsite team members were aided by an additional 10 remote team members, which allowed for the data processing rate to keep pace with the data upload rate. Although a detailed analysis of the fireball, recovery, and the meteorite itself (see Fig. 3) will be the subjects of a forthcoming study, this discovery cements the capability of this enhanced methodology for recovering meteorites.

**Success Within a Failure: Close Call Rockhole Meteorite Finds**

The fall site near Close Call Rockhole on the Nullarbor Plain was the first fall site surveyed and processed with the new methodology that is presented in this work, while it was still in development. A fall area of 1.3 $km^2$ ($2\sigma$) was calculated for the DN230523_02 fireball, and the area was fully surveyed in December 2024 at 1.8 mm $pixel^{-1}$, however the team was unable to complete the follow-up of all Stage 4 candidates due to time constraints. After the development of a more user-friendly interface for Stage 4, a second team visited the site and completed the search for the remaining candidates five months later (803 Stage 4 candidates total). Ultimately this search was unsuccessful in locating the orbital meteorite, however six weathered meteorite finds (see Fig. 3) unrelated to our targeted fall were recovered after being identified by the system, demonstrating the utility of this system to recover both pristine and weathered meteorites.

While the recovery of these older finds validated the new software developments and demonstrated that this updated methodology was an effective way to recover meteorite finds, the fact that we were not able to locate the fresh meteorite remains somewhat problematic. The meteorite could have been obscured by vegetation (Atriplex vesicaria and Cratystylis conocephala) that covered ~20% of the drone's view. It is also possible that the fireball and dark flight calculations were incorrect and the meteorite fell outside of the surveyed area. The remaining possibility is that the meteorite was either misclassified by the machine learning algorithm or mislabeled as a false positive by a user during Stage 1 or 2. Although determining the cause of failure is difficult in this individual case, systematic issues may become more apparent as we implement this methodology at multiple fall sites in distinct terrains.

**Failed Recoveries**

In November and December of 2023, prior to migrating our software to a remote server-hosted web application, we attempted to search for two meteorites near Kestrel Cavern and Bore Z, both in the western Nullarbor Plain. During these trips we used the YOLOv8 engine running on an onsite desktop to process the survey images.

In November 2023, we surveyed the meteorite fall site near Kestrel Cavern, two years and four months after a fireball event was observed in the area (July 2021, DN210727_01). During the trip, a rainstorm on the third day covered much of the area in standing water and prevented us from fully completing the survey. A follow-up trip 12 months later (November 2024) arrived on site to find that the heavy rains experienced the year before had spurred a bloom of nearly every plant type in the typically semi-arid region, including tall grasses that partially obscured a view of the ground from eye level. This made subsequent searching difficult, and due to the floral changes, the team could not finish the search, resulting in no recovered meteorites. We plan to re-survey this site in its entirety after at least another year has passed, and the vegetation thinned.

In December 2023, we conducted a survey near Bore Z where we predicted a small meteorite (< 50 g) had fallen a month earlier (event number: DN231024_01). Due to the small size of the meteorite and the time available, only the 1σ calculated fall area was searched (1.8 $km^2$) at 1.8 mm $pixel^{-1}$. The survey and data processing proceeded without issue. All Stage 1, 2, 3, and 4 candidates were evaluated using the onsite computer, via follow up drone, and in person, but the target meteorite was not found. By chance, a small meteorite find (approximately 3 mm in diameter) was recovered by a team member when evaluating Stage 4 candidates in person. This meteorite was not recognized by the machine learning algorithm, as it was below the detection limit of ~3 cm, and was only recovered due to the vigilant awareness of the team member. Later analysis via SEM revealed significant weathering, incompatible with a terrestrial age of less than one year of our target meteorite.

In July 2025 we enacted a final rehearsal to validate and test the new remotely hosted software tool. We chose a fall area near Booylgoo Springs, WA, outside of our typical search area in the Nullarbor plain and surveyed 2.6 $km^2$ of the large 11.5 $km^2$ 1σ search area at 2.6 mm $pixel^{-1}$. We were able to complete approximately half of the Stage 4 candidates before depleting our available time onsite. The fall site suffered from thick bush and tree cover, with 30-50% of the ground being obscured from the drone's perspective. The vegetation also prevented any vehicle movement off of the pre-existing dirt roads, which forced Stage 4 team members to spend significant amounts of time walking between the camp (where the internet terminal was located) and the search area, in order to reload new candidates on their mobile device. Although this trip was unsuccessful in recovering a meteorite, it was nonetheless useful in validating the software development completed thus far. This trip also revealed the future need for a native mobile application for Stage 4 that will be necessary to cache images to search at certain fall sites where internet connectivity is intermittent. This trip also highlighted the need to re-acquire training images as weather conditions (i.e. cloud cover) change, to more effectively eliminate false positives.

In March 2026 we searched at a fall site (DN150413_01) near Forrest Airport on the Nullarbor plain, with a surviving mass of ~0.1 kg. The 1σ search area (1.8 $km^2$) was surveyed at 1.8 mm $pixel^{-1}$, and candidate-false positive sorting was completed with 49,601 in Stage 1, 1,922 in Stage 2, and 1,033 in Stage 4, without successfully locating the meteorite. This site had significant vegetation due to recent rain, covering ~30 % of the area from the drone's perspective, and only the 1σ calculated search area, which could explain why the meteorite was not located. Unlike the Close Call Rockhole fall site, no meteorite finds were recovered. This could possibly be explained by the proximity to Forrest Airport, a settlement sometimes used in the past by meteorite hunters to search for meteorite finds, due to its dirt road access, and utility as a location to re-provision during longer searching campaigns.

A commonality among our failed meteorite recovery attempts appears to be a high proportion of the searching area covered in vegetation. This vegetation is typically in the form of trees or tall grasses that have the ability to extend over a meteorite on the ground. It is currently unclear how plant coverage could prevent meteorite recovery using our system in locations that are maintained by local inhabitants, such as cut crop fields or manicured residential land, like those in Europe or North America.

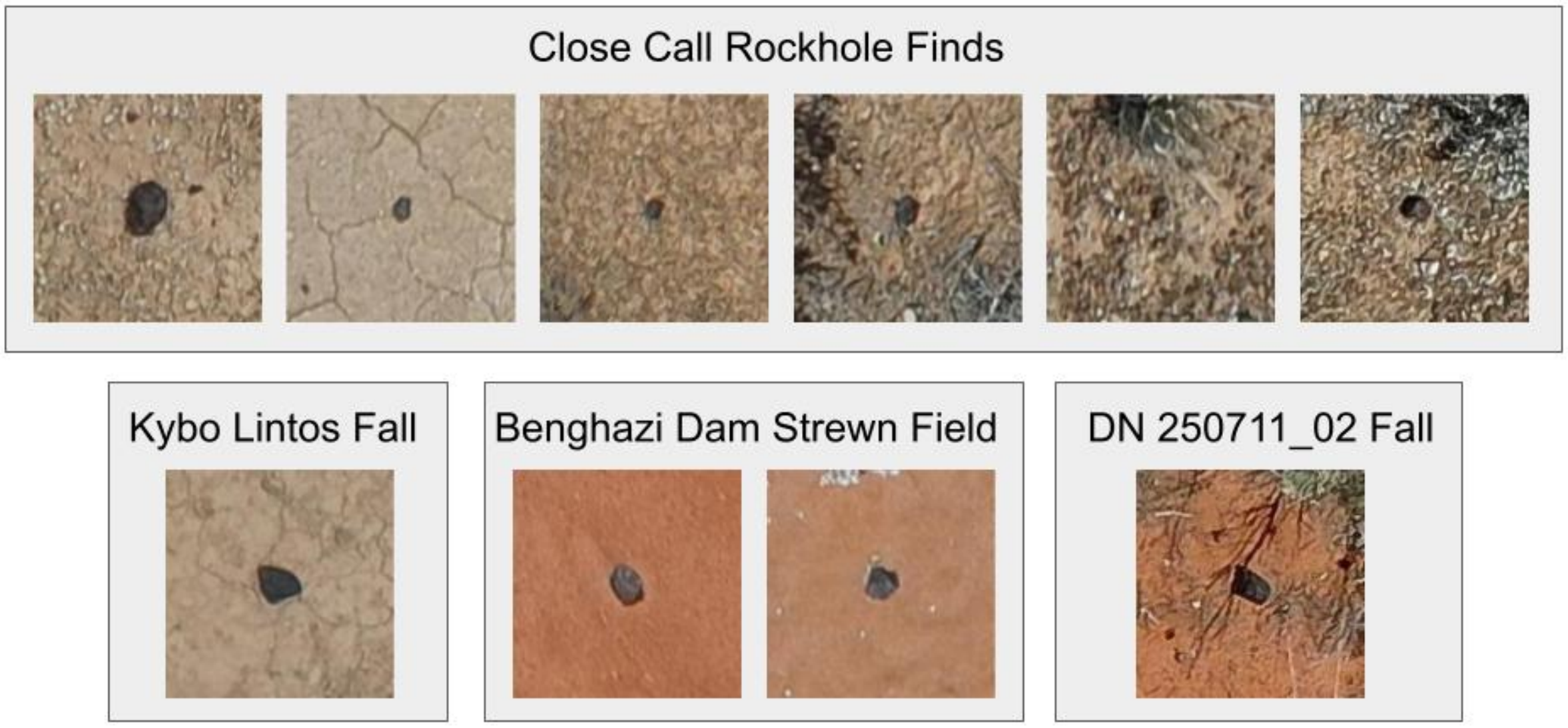

Fig. 3. Drone survey images of all our recovered meteorites. All images are the same scale, with each side equal to 30.5 cm.

## CONCLUSIONS AND FUTURE WORK

In this work, we present a comprehensive solution for using drones and machine learning to recover meteorite falls. The web application is accessible to the meteorite research community at https://find.gfo.rocks. In the subsections below, we detail some of our anticipated future work and improvements.

### Applicability to Other Terrains

Our original goal was to create an efficient tool for recovering meteorites on the Nullarbor plain in South and Western Australia (Anderson et al., 2020). The Nullarbor plain is a flat limestone plateau, mostly devoid of trees with usually less than 20% low shrub vegetation cover. This makes the terrain excellent for meteorite searching, inferior only to the conditions encountered in true deserts.

Testing the system in true deserts such as in Oman (Hofmann and Gnos, 2025), Atacama (Gattacceca et al., 2011; Sadka et al., 2025; 2026), or Antarctica (Cassidy et al., 1977) would be a logical next step, though we have also shown operational success in areas that are significantly more difficult to search than in the Nullarbor. We therefore plan to use the system in other areas such as Europe and North America, where freshly fallen meteorites are often found in cropped fields (Egal et al., 2025). Appropriately trained, the number of false positives in a cropped field could be very low. Additionally, using a drone would not flatten crops in the process of surveying, and would be a significant advantage compared to visual on-foot searching which requires five or more team members to walk in a line across the fall site.

**Future Work**

While the system presented in this work is significantly more efficient than traditional searching on foot, we list below some improvements and refinements we envision in the future:

- Image recognition software and machine learning models are constantly improving (e.g. Robinson et al., 2025), and we will therefore continue to update the machine learning engine to reduce the number of false positives that users must contend with.
- The Stage 4 interface will be improved as a native mobile device application, instead of the current web browser.
- Stage 3 of the searching process is not yet implemented in the current web application and could reduce the time spent by team members in the field eliminating false positives. Combining this step with the added capability of a multi-spectral sensor (see below) could further enhance the effectiveness of this step, though this addition could be limited by site-specific factors such as tree coverage and fall site total area.
- Multispectral imaging could enable greater discrimination between meteorites and surrounding materials by exploiting different reflectance properties across individual spectral bands. These differences can be enhanced by combining images from individual bands into false color composites, allowing meteorites to appear visually distinctive from their backgrounds as well as some terrestrial rocks (Lovelock et al., 2025). By automating this process, a user could be presented with an image showing meteorites in a distinctive color, supporting more efficient validation and reduction of false positives. Most multispectral sensors have a significantly lower spatial resolution than our survey drone's 48 MP camera, therefore such an approach would likely be best implemented on a follow up drone for Stage 3 searching.
- Thermal imaging also offers a potential enhancement for meteorite identification by exploiting differences in thermal inertia between meteorites and common terrestrial materials. Meteorites could be identified by their temperatures following exposure to sunlight, as they retain and release heat at different rates than terrestrial rocks and soils (Lovelock et al., 2025). This approach would require significant development before application in the field, including thermal modeling of meteorites to account for the effect of sample size and cooling influences, such as wind speed and ambient temperature on sample temperature.
- While thermal and multi-spectral imaging techniques could help to reduce the number of candidates and assist with searches conducted in more heavily vegetated terrain, especially in Stage 3, we do not anticipate these to fully replace RGB imaging in the survey phase, due to their significantly lower spatial imaging resolution.

**ACKNOWLEDGEMENTS**

This research was partially supported by an NPP (NASA Postdoctoral Program) appointment, administered by Oak Ridge Associated Universities (ORAU). This work was supported by software support resources awarded under the Astronomy Data and Computing Services (ADACS) Merit Allocation Program. ADACS is funded from the Astronomy National Collaborative Research Infrastructure Strategy (NCRIS) allocation provided by the Australian Government and managed by Astronomy Australia Limited (AAL). This research was supported by use of the Nectar Research Cloud, a collaborative Australian research platform supported by the NCRIS-funded Australian Research Data Commons (ARDC). Specifically, the Nectar Research Cloud enables us to run the web platform VM (*m3.xlarge*) and the GPU-enabled VM (*g2.large*). This work was supported by resources provided by the Pawsey Supercomputing Research Centre's Acacia Object Storage, with funding from the Australian Government and the Government of Western Australia. The Desert Fireball Network and Global Fireball Observatory programs have been funded by the Australian Research Council as part of the Australian Discovery Project scheme (grant Nos. DP170102529, DP200102073, and DP230100301), the Linkage Infrastructure, Equipment and Facilities scheme (grant No. LE170100106), and received institutional support from Curtin University. S.E.D. acknowledges the support of an Australian Government Research Training Program Scholarship.